\documentclass{iopjournal-modified}

\newcommand{\rbracket}[1]{\left({#1}\right)}
\newcommand{\sbracket}[1]{\left[{#1}\right]}

\newcommand{\at}[2]{\left.{#1}\right|_{#2}}

\newcommand{\T}{\mathcal{T}}
\newcommand{\de}{\partial}
\newcommand{\be}{\begin{equation}}
\newcommand{\ee}{\end{equation}}
\usepackage{mathtools}
\usepackage[bitstream-charter]{mathdesign}
\usepackage{tikz}
\usepackage{float}
\usetikzlibrary{arrows.meta}
\usepackage{rotating}
\usepackage{upgreek}
\usepackage{pdflscape}
\usepackage{tabularx}
\usepackage{array}
\usepackage{graphicx}
\usepackage{cite}
\usepackage{physics}
\usepackage[protrusion=true,expansion=true]{microtype}
\usepackage{times}
\usepackage{hyperref}
\usepackage{bookmark}
\hypersetup{
    pdftitle={Analytical results for thermal correlation enhancement after a trap quench},
    pdfauthor={Giorgio Li, Paola Ruggiero, Stefano Scopa},
    colorlinks=true,
    citecolor=blue,
    linkcolor=blue,
    urlcolor=blue
  }
\begin{document}
\articletype{Paper} 
\title{\bf Analytical results for thermal correlation enhancement after a trap quench}

\author{Giorgio Li$^{1,*}$\orcid{0009-0003-0196-6595}, Paola Ruggiero$^{1}$\orcid{0000-0002-0891-0477}, and Stefano Scopa$^2$\orcid{0000-0001-7638-8804}}
\affil{$^1$~Department of Mathematics, King's College London, Strand WC2R 2LS, United Kingdom}

\affil{$^2$~Laboratoire de Physique de l’\'Ecole Normale Sup\'erieure, CNRS,
ENS \& Universit\'e PSL, Sorbonne Universit\'e, Universit\'e Paris Cit\'e, 75005 Paris, France.}

\affil{$^*$Author to whom any correspondence should be addressed.}

\email{giorgio.1.li@kcl.ac.uk}

\begin{abstract}
We investigate the nonequilibrium evolution of density correlations in a thermal one-dimensional Bose gas following a trap quench. 
We focus on the Tonks--Girardeau limit and develop an analytical description based on a finite-temperature extension of Quantum Generalized Hydrodynamics. 
We benchmark our predictions against exact numerics for a quench from a double well to a harmonic trap and for a trap release, and assess their temperature range of validity. 
Our results capture the dynamical enhancement of correlations during the expansion stages, namely the crossover from thermally suppressed to power-law bulk correlations. 
For the trap release, the conformal distance associated with fixed bulk points decreases as $1/t$, so the correlations progressively probe a universal regime~$\sim1/x^2$ that is independent of temperature. 
We thus show that, in the hydrodynamic long-time regime, the bulk density correlations approach their zero-temperature equilibrium form independently of the initial temperature.
\end{abstract}

\section{Introduction}\label{sec:introduction}
A central challenge in modern condensed-matter physics is understanding the out-of-equilibrium dynamics of quantum many-body systems. One-dimensional systems provide a particularly favorable setting, where integrability and low-energy field theory offer complementary analytical descriptions. In particular, the Bethe ansatz provides a quasiparticle description of integrable models in terms of rapidities, whose ground-state distributions form Fermi-sea-like structures even for bosonic systems~\cite{Bethe_31,Takhtadzhan_Faddeev_79,Lieb_Liniger_63,Korepin_Bogoliubov_Izergin_93}, while, at least for gapless models, bosonization and Tomonaga-Luttinger liquid (TLL) theory describe their low-energy fluctuations~\cite{Haldane_81,Cazalilla_04,Giamarchi_03}. Building on these frameworks, a major development has been generalized hydrodynamics (GHD)~\cite{CastroAlvaredo_Doyon_Yoshimura_16,Bertini_Collura_Nardis_Fagotti_16}, which provides a large-scale description of integrable dynamics in terms of locally equilibrated quasiparticle distributions, even in the presence of inhomogeneities, where standard integrability breaks down.\\
These developments are closely connected with advances in experimental platforms, most notably ultra-cold atomic gases confined in one-dimensional traps~\cite{Kinoshita_Wenger_Weiss_04,Bloch_Dalibard_Zwerger_08,Schemmer_Bouchoule_Doyon_Dubail_19}, where the system can be prepared, driven out of equilibrium, and probed with high spatial and temporal resolution. A paradigmatic example is the quantum Newton's cradle~\cite{Kinoshita_Wenger_Weiss_06}, in which two counter-propagating atomic clouds undergo repeated collisions without appreciable thermalization over experimentally accessible times. Such experiments have provided direct tests of integrable dynamics and, more recently, of GHD itself~\cite{Schemmer_Bouchoule_Doyon_Dubail_19,malvania2021generalized,Horvath2026}. See Refs.~\cite{GHD_Review,Doyon_20,Alba2021,Essler_23,Guan_He_22,Bertini2021} for reviews on GHD and its connection to experimental realizations.\\
The evolution of the quasiparticle occupation alone, however, does not determine non-local quantum observables such as long-range correlations and entanglement. Inhomogeneous Luttinger-liquid and related conformal-field-theory methods have made it possible to describe such observables in trapped one-dimensional systems~\cite{Brun_Dubail_17,Brun_Dubail_18,scopa2020one,Bastianello_2020,Takacs_Zhang_Calabrese_Dubail_Rigol_Scopa_25}, by incorporating the spatial dependence generated by the trap into an effective low-energy field theory. Quantum Generalized Hydrodynamics (QGHD)~\cite{Ruggiero_Calabrese_Doyon_Dubail_20,Ruggiero_Calabrese_Doyon_Dubail_21} extends this construction to nonequilibrium hydrodynamic backgrounds. Restricting to zero-entropy states~\cite{Doyon_Dubail_Konik_Yoshimura_17}, the local quasiparticle occupation is characterized by a collection of Fermi seas whose boundaries define a Fermi contour in position-momentum space. QGHD quantizes the fluctuations of this contour, leading to a time-dependent spatially inhomogeneous, multicomponent Tomonaga-Luttinger theory. This framework has been used to obtain several analytical results for density correlations and entanglement following trap quenches of a Tonks-Girardeau gas, see e.g. Refs.~\cite{Ruggiero_Calabrese_Doyon_Dubail_21,Scopa_2023,scopa2023scaling}, as well as genuinely interacting Bose gases~\cite{Ruggiero_Calabrese_Doyon_Dubail_20,urilyon2024quantum,Scopa2022}.
The existing QGHD construction is naturally formulated around zero-entropy states, and hence is suited for the nonequilibrium dynamics starting from inhomogeneous ground states.  Experiments, however, are necessarily performed at finite temperature, where thermal and quantum fluctuations coexist. This raises the question of how non-local correlations evolve when an inhomogeneous one-dimensional quantum gas is driven out of equilibrium from a thermal initial state. The corresponding equilibrium problem has recently been addressed in Ref.~\cite{Takacs_Zhang_Calabrese_Dubail_Rigol_Scopa_25}, while the finite-temperature nonequilibrium dynamics remains much less explored. In this work, we address this problem for trap quenches and derive analytical predictions for the ensuing density-density correlations. We focus on the Lieb-Liniger gas~\cite{Lieb_Liniger_63,Korepin_Bogoliubov_Izergin_93} in the Tonks-Girardeau (TG) limit of infinitely strong contact repulsion~\cite{Girardeau_60}. In this limit, the inhomogeneous Tomonaga-Luttinger theory underlying QGHD can be recast, after an appropriate change of coordinates, in terms of a compact-boson conformal field theory~\cite{Dubail_Stephan_Viti_Calabrese_17}. 
Finite-temperature effects can then be incorporated by also compactifying the Euclidean-time direction of the same CFT.
This allows us to obtain a closed analytical expression for the time-dependent density-density correlations in terms of the evolving Fermi contour and
the equilibrium chiral-current correlator encoding the quantum density fluctuations.\\
We benchmark this finite-temperature QGHD prediction against exact numerics for different quench protocols. At sufficiently low temperatures, the result reproduces the zero-temperature prediction, while at higher temperatures it captures the thermal suppression of correlations in the initial state and their subsequent nonequilibrium evolution. We explicitly characterize the temperature range of validity of our formulas. Remarkably, during the evolution, correlations can become progressively closer to their zero-temperature values, and the finite-temperature QGHD prediction remains quantitatively accurate even beyond its expected low-temperature regime of validity.
This behaviour is reminiscent of expansion-based cooling protocols used in ultra-cold gases, where the expansion of the atomic cloud is combined with a subsequent pulse to reduce its momentum spread, see e.g. Refs.~\cite{Ammann_Christensen_97,Myrskog_Fox_Moon_Kim_Steinberg_00,dupays2021delta,Dupays_Yang_delCampo_23}. Here, the finite-temperature QGHD framework provides analytical access directly to the correlation dynamics and allows us to establish the long-time convergence toward the zero-temperature equilibrium correlation after a trap release. 
Indeed, the long-time state qualitatively differs from the corresponding zero-temperature equilibrium configuration only by a local momentum boost, to which density correlations are insensitive due to Galilean invariance.
The mechanism is controlled by the long-time evolution of the insertion points along the Fermi contour: their separations, expressed in co-moving coordinates along the contour, shrink over time such that the universal singularity of the chiral-current correlator eventually dominates on short mesoscopic scales. 
Since this singular contribution is temperature independent and coincides with the zero-temperature correlator at small mesoscopic distances, regular temperature-dependent contributions are progressively suppressed.
This provides a quantitative extension, for trap-release dynamics, of the correlation-enhancement mechanism previously observed in the melting of a domain-wall state~\cite{DeAngelis_DeNardis_Scopa_24}.\\

The paper is organized as follows. In Sec.~\ref{sec:setup} we introduce the model, specialize it to the TG limit, and define the quench protocol together with the connected density-density correlation which is our central observable. In Sec.~\ref{sec:qghd} we set up the finite-temperature QGHD framework for the quench protocol introduced above. The framework combines two ingredients: a classical, hydrodynamic evolution of the quasiparticle occupation in phase space, which fixes the time evolution of a Fermi contour, and an effective field theory for the fluctuations living on that contour. Combining them, we show that the post-quench density-density correlation reduces to an equilibrium correlator of a chiral field along the Fermi contour, so that the entire effect of the quench is encoded in how the contour evolves in time. In Sec.~\ref{sec:luttinger} we evaluate this correlator using inhomogeneous TLL theory. We show that a suitable change of coordinates absorbs the spatial inhomogeneity and maps the theory onto a CFT in curved space. We then apply these results to two quench protocols. In Sec.~\ref{sec:double_well_results} we consider a quench from a double well to a harmonic trap and benchmark our prediction against exact numerics across the different temperature regimes. In Sec.~\ref{sec:trap_release} we consider a trap release, for which we compute the long-time asymptotics and show analytically that density-density correlations converge to their zero-temperature equilibrium form. We conclude in Sec.~\ref{sec:conclusion}.

\section{Model, quench protocol, and observables}\label{sec:setup}
We consider a one-dimensional gas of bosons with repulsive contact interactions of strength $g>0$ in a confining potential $V_0(x)$. Its Hamiltonian is
\begin{equation}\label{eq:LL-model}
    \hat H_{0;\mu}(g) = \int_\mathbb{R} \dd x\ \left[ \frac12\partial_x\hat{\psi}^\dagger\partial_x\hat{\psi} + \big(V_0(x)-\mu\big)\hat\rho(x) + \frac{g}{2}(\hat{\psi}^\dagger)^2\hat{\psi}^2 \right].
\end{equation}
Here $\hat{\psi}^\dagger(x),\hat{\psi}(x)$ are the bosonic creation and annihilation operators at position $x$, with the canonical commutation relations $\left[ \hat{\psi}(x), \hat{\psi}^{\dagger}(x') \right] = \delta(x-x')$, and $\hat\rho(x)=\hat\psi^\dagger(x)\hat\psi(x)$ is the density operator. We work in the grand-canonical ensemble with chemical potential $\mu$, so that the local density is set by the effective chemical potential $\mu-V_0(x)$.
\\
\indent
In the absence of a confining potential, the model reduces to the Lieb-Liniger gas, which is exactly solvable by coordinate Bethe ansatz~\cite{Korepin_Bogoliubov_Izergin_93, Takahashi1999}. A nonzero trap, $V_0\neq 0$, breaks this integrability.  Nevertheless, by combining the thermodynamic Bethe ansatz~\cite{Yang_Yang_69} with the local density approximation, an accurate description of local observables and of their dynamics via GHD can be achieved, see e.g. Refs.~\cite{Doyon2017,Caux_Doyon_Dubail_Konik_Yoshimura_19,Bastianello2019,Bastianello2020,Doyon_Dubail_Konik_Yoshimura_17,Scopa2022generalized,scopa2026absence,Bastianello2020therm}.
Importantly, however, this local description does not extend to non-local observables such as the two-point density-density correlation function $\langle\hat\rho(x)\hat\rho(y)\rangle$, which is the main focus of this work. \\
\indent
In the following, we will consider the strong-repulsion or TG limit $g\to\infty$ of the model~\eqref{eq:LL-model}, where simplifications occur. In fact, it has long been known that in this limit the bosonic Hamiltonian~\eqref{eq:LL-model} becomes quadratic,
\be\label{eq:TG-H}
\hat H_{0;\mu}(\infty)= \int_\mathbb{R} \dd x\ \left[ \frac12\partial_x\hat{\psi}^\dagger_F\partial_x\hat{\psi}_F + \big(V_0(x)-\mu\big)\hat\rho(x)\right],
\ee
and can be expressed in terms of spinless fermionic operators $\hat \psi_F(x)$, $\hat{\psi}_F^\dagger(x)$ via the Jordan-Wigner transformation,
\begin{equation}
    \hat{\psi}^\dagger_F(x) = e^{i\pi \int_{y<x}\; \hat{\rho}(y) \dd y}\hat{\psi}^{\dagger}(x),
\end{equation}
with $\{\hat{\psi}_F(x),\hat{\psi}^{\dagger}_F(x')\}=\delta(x-x')$. As discussed below, our calculation of the correlation dynamics relies on a TLL description whose spatial inhomogeneity can be absorbed into a conformal coordinate transformation only at the free-fermion point $g\to\infty$~\cite{Dubail_Stephan_Viti_Calabrese_17,Brun_Dubail_17}. An analogous construction exists at finite interaction~\cite{scopa2020one,Bastianello_2020,Takacs_Zhang_Calabrese_Dubail_Rigol_Scopa_25}, but it generally requires numerics and does not yield closed analytical results.
\\
We consider the following protocol. At $t=0$, the system is prepared in the thermal state
\begin{equation}
    \hat\varrho_{\beta}(0) = \frac{\exp\left(-\beta \hat H_{0;\mu}(\infty)\right)}{\Tr\sbracket{\exp\left(-\beta \hat H_{0;\mu}(\infty)\right)}}
\end{equation}
with inverse temperature $\beta=1/T$ and chemical potential $\mu=\mu(N,\beta)$ tuned so that the average particle number $\langle \hat N\rangle_\beta =N$ is fixed at all temperatures considered.

At $t=0^+$ the confining potential is suddenly varied, $V_0(x)\to V_1(x)$, which quenches the Hamiltonian $\hat H_{0;\mu}\to \hat H_{1;\mu}$. For $t>0$, we study the resulting nonequilibrium dynamics.
Throughout the rest of the paper, we use the notation
\be
\langle\cdot\rangle_{t;\beta}=\Tr\left[\hat\varrho_{\beta}(t)\ \cdot\right],
\ee
where $\hat\varrho_{\beta}(t)$ is specified by the post-quench unitary dynamics, $\dd\hat\varrho_{\beta}(t)=-i[\hat H_{1;\mu}(\infty),\hat\varrho_{\beta}(t)] \dd t$.
We study the post-quench evolution of the connected density-density correlation
$\langle\delta\hat\rho(x)\delta\hat\rho(y)\rangle_{t;\beta}$, with $\delta\hat\rho=\hat\rho-\langle\hat\rho\rangle_{t;\beta}$. For such density correlations, our results below apply equally to impenetrable bosons and to fermions.

Note that, even though the many-body Hamiltonian is Gaussian in the TG limit, standard free-fermion methods do not generally yield closed analytical results for the correlation dynamics, especially for arbitrary trap quenches. Exact results are available only for special confining potentials~\cite{Dean2016,Dean2018}, most notably harmonic traps~\cite{minguzzi2005exact,minguzzi2022strongly,Dean2015,Scopa2018,Ruggiero_Brun_Dubail_19,Grela2017,Rigol_Muramatsu_05}. A systematic route to more general inhomogeneous settings is instead provided by QGHD~\cite{Ruggiero_Calabrese_Doyon_Dubail_20}, recently extended to finite temperature in the equilibrium regime~\cite{Takacs_Zhang_Calabrese_Dubail_Rigol_Scopa_25}. In the following section we revisit this framework and extend it to the quench dynamics considered here.

\section{Finite-temperature Quantum Generalized Hydrodynamics framework}\label{sec:qghd}
The main strategy underlying the QGHD framework is to separate the hydrodynamic evolution of the quasiparticle distribution from the effective description of the relevant long-range fluctuations. As discussed below, the first task is implemented through a local-density approximation and yields a phase-space dynamics. At this stage, the evolution is local and classical, and is therefore not suited to describing non-local quantum correlations. The latter are then reintroduced through an effective CFT, analogous to a finite-temperature TLL model but accounting for the density modulations induced by the trap.

\subsection{Phase-space evolution}
We first introduce the Wigner function~\cite{Wigner_32, Hillery_OConnell_Scully_Wigner_84,Cahill_Glauber_69}
\begin{equation}\label{eq:Wigner-func}
    W_{t;\beta}(x,p) = \int_\mathbb{R} e^{ipy}\expval{\hat{\psi}^\dagger_F\rbracket{\frac{x+y}{2}}\hat{\psi}_F\rbracket{\frac{x-y}{2}}}_{t;\beta} \dd y,
\end{equation}
which carries the semiclassical interpretation of a quasi-probability of finding a particle with position $x$ and momentum $p$. Its dynamics is governed by the Moyal equation~\cite{Moyal_49} and, to leading order in $\de_x$ and $\de_p$ derivatives, reduces to the classical Liouville equation
\be\label{eq:Wigner-evo}
\de_t W_{t;\beta} +p \de_x W_{t;\beta} = \de_x V_1 \de_p W_{t;\beta}.
\ee
Note that Eq.~\eqref{eq:Wigner-evo} is exact when the post-quench potential is harmonic, linear, or switched off. Higher-order corrections have recently been considered in Refs.~\cite{Dean2018,Fagotti_17,Fagotti_20}.\\
\indent
Importantly, we shall not work with the full Wigner function in Eq.~\eqref{eq:Wigner-func}. Instead, we consider the non-fluctuating coarse-grained occupation obtained within the local density approximation,
\be
 n_{t;\beta}(x,p) =\frac{1}{\Delta}\int_{x-\frac{\Delta}{2}}^{x+\frac{\Delta}{2}} \dd z\, W_{t;\beta}(z,p),
\ee
with $\Delta$ a mesoscopic fluid cell over which the potential $V(x)$ is approximately constant, $V(y)\approx V(x)$ for $x-\Delta/2\leq y\leq x+\Delta/2$. Under the separation of scales hypothesis~\cite{Doyon_20}, we shall assume that $\Delta$ is large enough for local thermodynamics to apply, while remaining small compared with the hydrodynamic scales. This quantity takes values $n_{t;\beta}\in[0,1]$ and corresponds to the non-interacting limit of thermodynamic Bethe ansatz occupation functions, see e.g. Refs.~\cite{Doyon_20,Alba2021}. By construction, it does not carry quantum fluctuations. Its evolution,
\be\label{eq:phase-space-evo}
\de_t n_{t;\beta}+ p\de_x n_{t;\beta}=\de_xV_1\de_p n_{t;\beta},
\ee
describes a classical Hamiltonian flow in phase space. At time $t=0$, the occupation function is the one of a thermal state and reads
\begin{equation}\label{eq:FD-distr}
    n_{0;\beta}(x,p) = \frac{1}{1+e^{\beta\sbracket{\frac{p^2}{2}+V_0(x)-\mu}}}.
\end{equation}
We mention that the possibility of constructing and evolving a fluctuating semiclassical Wigner function has been explored in Refs.~\cite{bettelheim2011universal,bettelheim2012quantum}, and more recently in Ref.~\cite{urilyon2024quantum} for interacting models. Our approach, however, is different and we only consider the occupation function $n_{t;\beta}(x,p)$ introduced above.

\subsection{Effective field theory description}
\label{sec:effective_theory}
We then supplement the phase-space dynamics in Eq.~\eqref{eq:phase-space-evo} with an effective field-theory description. As in standard field theory, such a description encodes both thermal and quantum fluctuations in terms of low-energy operators defined on a background, which must first be specified. In the QGHD prescription~\cite{Ruggiero_Calabrese_Doyon_Dubail_20}, the natural reference background is set by the zero-temperature phase-space dynamics described by $n_{t;\beta\to\infty}$ with chemical potential $\mu(N,\beta\to\infty)$. Unless otherwise specified, we shall denote $n_t \equiv n_{t;\infty}$ and $\mu_\infty \equiv \mu(N,\beta\to\infty)$.
We shall retain this zero-temperature contour as the reference background also at finite temperature. As in the standard finite-temperature TLL description, this construction is expected to apply as long as thermal broadening does not wash out the underlying local Fermi points. We return to this regime of validity in Sec.~\ref{sec:double_well_results}.
\\

At time $t=0$, the zero-temperature occupation is $n_0=\Theta(p_F(x)-|p|)\in\{0,1\}$, with $\Theta(\cdot)$ the Heaviside step function. In other words, it encodes the occupations within the local Fermi seas $[-p_F(x),p_F(x)]$, with
\begin{equation}\label{eq:fermi_point}
    p_F(x) = \sqrt{2\sbracket{\mu_\infty-V_0(x)}}
\end{equation}
where $p_F(x)\geq 0$. In general, the resulting confining region for the gas will be given by $x\in\bigcup_{j=1}^{\kappa} [x_{2j-1},x_{2j}]$, with $x_{j}$  solutions of $V_0(x_{j})-\mu_\infty=0$. For simplicity, we consider $V_0$ such that the gas is initially confined in the region $x\in[-R,R]$, with a straightforward generalization to the case $\kappa>1$ and to asymmetric intervals.\\

Since $n_0$ is a collection of non-fluctuating Fermi seas, it has zero entropy. This zero-entropy condition is preserved by the Euler evolution \eqref{eq:phase-space-evo}, as noted in Ref.~\cite{Doyon_Dubail_Konik_Yoshimura_17}. It is therefore convenient to describe $n_t$ solely through its boundary $\Gamma_t$ in the $(x,p)$ plane, such that
\begin{equation}
    n_t(x,p) = \begin{cases}
        1,\quad \text{if $(x,p)$ is inside $\Gamma_t$;}\\
        0, \quad \text{otherwise.}
    \end{cases}
\end{equation}
For fixed time $t$, $\Gamma_t$ is obtained by evolving each point of the initial contour according to
\be\label{eq:hamilton_eom}
\dv{t}\mqty(x_t\\p_t)=\mqty(p_t\\-\partial_xV_1(x_t)),
\ee
with initial contour
\be
\Gamma_0=\Big\{(x_0,p_0)\ : p_0^2=p_F^2(x_0)\Big\}.
\ee
Crucially, $\Gamma_t$ defines the hydrodynamic background for the effective field theory discussed below.\\

We now introduce fluctuating fields on top of this background. At a fixed space-time position $(x;t)$, we denote by $\{p_a(x;t)\}_{a=1}^{2Q_t(x)}$ the ordered set of local Fermi points such that, for each $a \in \{1, \cdots,  2 Q_t (x) \}$, it holds $\big(x_t\equiv x, p_t\equiv p_a(x;t)\big)\in \Gamma_t$. We then consider fluctuations of the local Fermi point $p_a(x;t) \to p_a(x;t) +\delta p_a(x;t)$. For the TG gas \eqref{eq:TG-H}, such momentum fluctuations are related to chiral density fluctuations via
\be\label{eq:dens-fluc}
\delta\rho_a(x;t)=(-1)^a\frac{\delta p_a(x;t)}{2\pi},
\ee
while an analogous construction applies to interacting integrable systems upon parametrizing the Fermi contour in terms of physical momentum rather than rapidity.
For instance, if $Q_t(x)=1$, i.e., there are locally only two Fermi points, Eq.~\eqref{eq:dens-fluc} simply states that the density fluctuation is
\be
\delta\rho(x;t)=\sum_{a=1}^{2}\delta\rho_a(x;t)=\frac{\delta p_2(x;t)-\delta p_1(x;t)}{2\pi}.
\ee
At this stage, the fluctuations in Eq.~\eqref{eq:dens-fluc} are still classical variables.

The QGHD quantization prescription promotes these chiral density fluctuations, associated with local deformations of each Fermi edge, to quantum fields, $\delta\rho_a\to\delta\hat\rho_a$. Their dynamics is governed by the effective quadratic Hamiltonian~\cite{Ruggiero_Calabrese_Doyon_Dubail_21}
\be\label{eq:H-eff}
\hat{H}_\text{eff}[\Gamma_t]=\pi\int_\mathbb{R} \dd x\; \sum_{a=1}^{2Q_t(x)}(-1)^a v_a(x;t)\ \big(\delta\hat\rho_a(x;t)\big)^2,
\ee
where $v_a(x;t)\equiv p_a(x;t)$ is the local chiral velocity. The quantum fields $\delta\hat\rho_a$ obey the $U(1)$ current algebra inherited from the canonical quantization of Fermi-contour deformations~\cite{Giamarchi_03}
\begin{equation} \label{U1_KM_algebra}
    \sbracket{\delta\hat{\rho}_a(x;t),\delta\hat{\rho}_b(y;t)} = \frac{(-1)^a}{2\pi i}\delta_{ab}\delta'(x-y).
\end{equation}
Together, \eqref{eq:H-eff} and \eqref{U1_KM_algebra}
describe an inhomogeneous time-dependent multi-component TLL on the hydrodynamic background $\Gamma_t$.

As already noted e.g. in Refs.~\cite{scopa2021exact,Ruggiero_Calabrese_Doyon_Dubail_21,Scopa_2023}, the effective Hamiltonian \eqref{eq:H-eff} admits a natural description in terms of a one-dimensional coordinate $s$ along the contour $\Gamma_t$, rather than in terms of the real-space coordinate $x$. For convenience, we take $s\in[0,2\pi)$ such that
\begin{equation}
\Gamma_t=\left\{\big(x_t(s),p_t(s)\big):s\in[0,2\pi)\right\}.
\end{equation}
Choosing the orientation of $s$ consistently with the ordering of the Fermi points, one defines
\begin{equation} \label{U1_KM_algebra_s}
    \delta\hat{\rho}_t(s) := \abs{\dv{x_t(s)}{s}}\delta \hat{\rho}_a(x_t(s);t);\qquad [\delta\hat\rho_t(s),\delta\hat\rho_t(s')]=\frac{1}{2\pi i}\delta'(s-s'),
\end{equation}
where the index $a$ is chosen such that $p_t(s)=p_a(x_t(s);t)$. 
Note that the absolute value of the Jacobian ensures the proper transformation of the density measure, \(\delta\hat\rho_t(s)\,ds=\delta\hat\rho_a(x)\,|dx|\), while the branch orientation is encoded in the choice of \(s\).
The corresponding effective Hamiltonian is
\be\label{eq:H-eff-s}
\hat H_\text{eff}[\Gamma_t]=\pi\int_0^{2\pi} \dd s\; \left(\frac{\dd x_t(s)}{\dd s}\right)^{-1} p_t(s)\ (\delta\hat\rho_t(s))^2.
\ee
In terms of the contour coordinate $s$, one finds the equation of motion~\cite{Ruggiero_Calabrese_Doyon_Dubail_21}
\be\label{eq:time_evolution_s}
\frac{\dd}{\dd t}\delta\hat\rho_t(s)=i[\hat H_\text{eff}[\Gamma_t],\delta\hat\rho_t(s)] +\de_s\left(\left(\frac{\dd x_t(s)}{\dd s}\right)^{-1} p_t(s)\ \delta\hat\rho_t(s)\right)=0.
\ee
This shows that the evolution of the density fluctuations is entirely induced by the time dependence of the hydrodynamic background. In terms of the co-moving coordinate $s$, these fluctuations therefore do not evolve. This property reflects the conformal structure of the TG gas (see discussion in Sec.~\ref{sec:luttinger}), for which the deformation of the background can be absorbed into the appropriate coordinate transformation. Note that, instead, for generic interacting systems the right-hand side of Eq.~\eqref{eq:time_evolution_s} is not always zero, see Ref.~\cite{Ruggiero_Calabrese_Doyon_Dubail_20}.
This observation also implies that, in the $s$ coordinate, the effective theory of fluctuations can, de facto, be studied at $t=0$. The effect of the quench dynamics is entirely transferred to the insertion points $s_a(x;t)\equiv s\big(x,p_a(x;t)\big)$ of the low-energy operators through their time-dependent relation to the real-space coordinates.

We now make this relation explicit.
As per \eqref{U1_KM_algebra_s}, also in the contour coordinate \(s\) the density fluctuations obey a \(U(1)\) current algebra. The latter admits the standard bosonized representation in terms of a compact chiral bosonic field \(\hat\phi(s)\)
\be\label{eq:chiral-boson}
\delta\hat\rho_t(s)=\frac{1}{2\pi}\de_s\hat\phi(s),
\ee
with $\hat\phi(s+2\pi)=\hat\phi(s)+2\pi \delta \hat N$, where $\delta \hat N\equiv \hat N-\langle \hat N\rangle_\beta$ denotes the total particle-number fluctuation
\be
\delta \hat N=\int_0^{2\pi}\dd s\,\delta\hat\rho_t(s).
\ee

The (long-wavelength) density fluctuation at the space-time point $(x;t)$ is then obtained by summing over all the local Fermi points,
\be\label{eq:dens-chiral-boson}
\delta\hat\rho(x;t)=\sum_{a=1}^{2Q_t(x)}\abs{\dv{s_a(x;t)}{x}}\delta\hat\rho_t\big(s_a(x;t)\big)
=\frac{1}{2\pi}\sum_{a=1}^{2Q_t(x)}\abs{\dv{s_a(x;t)}{x}}\;
\de_s\hat\phi(s)\Big\vert_{s=s_a(x;t)}.
\ee
Consequently, for non-coincident points, the connected density-density correlation can be written as
\be\label{eq:dens-dens-chiral-boson}
\langle\delta\hat\rho(x)\delta\hat\rho(y)\rangle_{t;\beta}
=\frac{1}{(2\pi)^2}
\sum_{a=1}^{2Q_t(x)}
\sum_{b=1}^{2Q_t(y)}
\abs{\dv{s_a(x;t)}{x}}
\abs{\dv{s_b(y;t)}{y}}\times
\left\langle
\de_s\hat\phi(s)\de_{s'}\hat\phi(s')
\right\rangle_{\Gamma;\beta}
\Big\vert_{\substack{s=s_a(x;t)\\s'=s_b(y;t)}}.
\ee
      
All dependence on the quench dynamics is therefore encoded in the time-dependent insertion points $s_a(x;t)$ and in the associated Jacobians. The notation $\expval{\ldots}_{\Gamma;\beta}$ indicates that the density fluctuations are equivalent (at leading order) to excitations living solely on the contour $\Gamma$. Indeed, Eq.~\eqref{eq:dens-dens-chiral-boson} can be read as a change of basis: from real space, on the left-hand side, to a basis supported entirely on $\Gamma$. As we anticipate here and make precise below, these excitations on the contour are described by an inhomogeneous Gaussian free field (see Ref.~\cite{Brun_Dubail_18} for details). For the TG gas, the choice of the coordinate $s$ can be made such that
this inhomogeneous theory is mapped onto a standard compact-boson CFT~\cite{Dubail_Stephan_Viti_Calabrese_17}. The finite-temperature effects are incorporated by compactifying the Euclidean-time direction of the same CFT.

\section{Inhomogeneous Gaussian free field at finite temperature}\label{sec:luttinger}
In the previous section, we showed that the calculation of $\langle\delta\hat\rho(x)\delta\hat\rho(y)\rangle_{t;\beta}$ reduces to evaluating the equilibrium chiral-current correlation function $\langle\de_s\hat\phi(s)\de_{s'}\hat\phi(s')\rangle_{\Gamma;\beta}$ on the Fermi contour. 
We now recast the effective theory of Sec.~\ref{sec:effective_theory}, i.e., the free chiral boson with a spatially varying velocity of Eq. \eqref{eq:H-eff}, in Euclidean path-integral form, which allows us to evaluate finite-temperature correlation functions by exploiting the conformal structure of the theory.
Following Refs.~\cite{Brun_Dubail_17,Brun_Dubail_18,scopa2020one,Takacs_Zhang_Calabrese_Dubail_Rigol_Scopa_25,Bastianello_2020}, this takes the form of an inhomogeneous Gaussian free field, whose Euclidean action at inverse temperature \(\beta\) is given by
\begin{equation}\label{eq:inhomogeneous_luttinger_action}
    S [h]=\frac{1}{8\pi}\int_0^\beta \dd\tau \int_{-R}^R
    \frac{\dd x}{K(x)}\sbracket{
        p(x)(\partial_x h)^2
        +\frac{1}{p(x)}(\partial_\tau h)^2}.
\end{equation}
where $h=h(x,\tau)$ is the height field entering the Euclidean path integral. The coordinate $\tau\in[0,\beta]$ denotes imaginary time, with the periodic identification $\tau\equiv\tau+\beta$ implementing the thermal trace. For the TG gas the \emph{Luttinger parameter} $K(x)=1$~\cite{Brun_Dubail_17}, so the local sound velocity $v(x)\equiv p(x)$ is
the only source of spatial inhomogeneity. This inhomogeneity can be absorbed by introducing the stretched
coordinates~\cite{Dubail_Stephan_Viti_Calabrese_17}
\begin{equation}\label{eq:s_mapping}
    s(x)=\frac{\pi}{\T}\int_{-R}^{x}\frac{\dd y}{p(y)},
    \qquad
    \T\equiv\int_{-R}^{R}\frac{\dd y}{p(y)},
\end{equation}
which maps $x\in[-R,R]$ onto $s\in[0,\pi]$. This is the same coordinate used
to parametrize the upper branch of the initial Fermi contour $\Gamma_0$. It extends to $s\in[0,2\pi)$ and parametrizes the complete contour after chiral unfolding, see discussion below. Rescaling imaginary time by the same factor,
\begin{equation}\label{eq:tau_rescaling}
    q=\frac{\pi}{\T}\,\tau,
\end{equation}
the action~\eqref{eq:inhomogeneous_luttinger_action} takes the homogeneous form
\begin{equation}\label{eq:homogeneous_luttinger_action}
    S = \frac{1}{8\pi}\int_{0}^{\beta_{\text{eff}}}\dd q
    \int_{0}^{\pi} \dd s
    \sbracket{(\de_s h)^2+(\de_q h)^2},
    \qquad
    \beta_{\text{eff}}\equiv\frac{\pi\beta}{\T}.
\end{equation}
All spatial dependence has thus been absorbed into the coordinate
transformation, while the physical inverse temperature enters the effective theory through the dimensionless combination $\beta_{\text{eff}}=\pi\beta/\T$.\\

The action~\eqref{eq:homogeneous_luttinger_action} is that of a homogeneous massless free boson CFT. 
On shell, its equation of motion implies the standard decomposition into holomorphic and anti-holomorphic components,
\begin{equation}\label{eq:height-chiral-decomposition}
h(s,q)=\varphi(s+iq)+\bar{\varphi}(s-iq).
\end{equation}
Since the gas is confined to $[-R,R]$ with no particle current at the boundaries of the trap, the height field satisfies Neumann boundary conditions,
\begin{equation}
    \at{\partial_s h}{s=0,\pi}=0,
\end{equation}
which relates the holomorphic and anti-holomorphic components at the boundaries $s=0,\pi$. The coordinate on the interval $s\in[0,\pi]$ can therefore be unfolded onto the circle $s\in[0,2\pi)$ by introducing the single chiral field
\begin{equation}\label{eq:unfolded-chiral-field}
    \phi(z)=
    \begin{cases}
        \varphi(z), & 0\leq \Re(z)\leq\pi,\\[1mm]
        \bar{\varphi}(2\pi-z), & \pi\leq \Re(z)<2\pi;
    \end{cases}
\end{equation}
with complex coordinate $z\equiv s+i q$. In terms of this chiral field, the height field on the original interval is
\begin{equation}\label{eq:height-unfolded-relation}
h(s,q)=\phi(s+iq)+\phi(2\pi-s+iq),\qquad s\in[0,\pi],
\end{equation}
up to an additive constant mode, which does not contribute to the correlations considered below. The particle-number zero mode $\delta N$ is instead encoded in the monodromy of the chiral field,
\begin{equation}\label{eq:chiral-monodromy}
    \phi(z+2\pi)=\phi(z)+2\pi\delta N.
\end{equation}

The Gaussian field theory above is the path-integral representation of the effective field theory on the contour. The unfolded $\phi$ corresponds to the chiral quantum field $\hat{\phi}$ entering Eq.~\eqref{eq:dens-dens-chiral-boson}. Throughout we use hats to denote quantum operators, dropping them only in the path-integral representation, where the fields are integration variables. Consequently,
\be\label{eq:CFT-mapping}
\left\langle
\de_s\hat\phi(s)\de_{s'}\hat\phi(s')
\right\rangle_{\Gamma;\beta}
\ =
\left\langle
\de\phi(z)\de\phi(z')
\right\rangle_{\text{CFT};\beta}\Big\vert_{\substack{\Re(z)=s \\ \Re(z')=s'}}
\ee
and this allows us to express Eq.~\eqref{eq:dens-dens-chiral-boson} entirely in terms of standard CFT correlation functions.

In terms of the complex coordinate $z\equiv s+iq$, the unfolded Euclidean geometry is specified by the identifications
\begin{equation} \label{eq:torusBC}
z\equiv z+2\pi, \qquad z\equiv z+i\beta_{\text{eff}}.
\end{equation}
These two periodic identifications define a torus with modular parameter
\begin{equation}\label{eq:Omega}
    \Omega=\frac{i\beta_{\text{eff}}}{2\pi}
    =\frac{i\beta}{2\T}.
\end{equation}
On this geometry, the two-point correlation is known and reads as~\cite{DiFrancesco_CFT}
\begin{equation}\label{eq:phiphi_torus}
    \expval{\phi(z)\phi(0)}_{\text{CFT};\beta}
    =-\log\sbracket{
        \frac{\theta_1(z/2\,|\,\Omega)}
        {\partial_z\theta_1(0\,|\,\Omega)}}
        +\frac{(\Im z)^2}{\pi \Im \Omega},
\end{equation}
and, in particular,
\begin{equation}\label{eq:dphidphi_torus}
    \expval{\partial\phi(z)\partial\phi(0)}_{\text{CFT};\beta} = \partial_z^2\log\theta_1(z/2|\Omega)+\frac{1}{2\pi\Im\Omega},             
\end{equation}
entering the finite-temperature QGHD expression for the density-density correlation in Eq.~\eqref{eq:dens-dens-chiral-boson}. Here $\theta_1$ is the Jacobi theta function~\cite{math_handbook}
\begin{equation}
    \theta_1(z|\Omega) = 2k^{1/4}\sum_{n=0}^{\infty} (-1)^n k^{n^2+n}\sin\sbracket{\rbracket{2n+1}z}, \quad k = e^{i\pi\Omega}.
\end{equation}
In our convention, $\phi(s)$ contains the charge zero mode $\delta N$, whose contribution to the density-density correlation is encoded in the additive constant of Eq.~\eqref{eq:dphidphi_torus}. Namely,
\be
\langle (\delta N)^2\rangle_{\text{CFT};\beta}
=
\frac{1}{2\pi\Im\Omega}
=
\frac{\T}{\pi \beta}.
\ee
As expected, at finite temperature this zero mode has a nonvanishing variance and therefore contributes to the density correlations. Conversely, in the zero-temperature limit, one finds $\langle (\delta N)^2\rangle_{\text{CFT};\beta\to\infty}\to 0$, since the ground state has a fixed particle number, without fluctuations. By noticing this, and that
\be
\lim_{\beta\to\infty}\partial_z^2\log\theta_1\left(\frac{z}{2} \Big|\frac{i\beta}{2{\T}}\right)=\frac{-1}{4\sin^2(z/2)},
\ee
one can recover the zero-temperature QGHD results of Ref.~\cite{Ruggiero_Calabrese_Doyon_Dubail_21}.
\section{Discussion of the results}\label{sec:double_well_results}
We now compare the finite-temperature QGHD result for the density-density correlation derived in Sec.~\ref{sec:qghd} and Sec.~\ref{sec:luttinger} with exact numerical data for the quench dynamics. 
While our framework gives access to correlations in any trap potential, in order to illustrate our findings, we consider a quench from a double-well to a harmonic potential. The latter shares similar dynamics with the well-known quantum Newton's cradle protocol, and the original setup was first experimentally realized in Ref.~\cite{Kinoshita_Wenger_Weiss_06}. Since then, the quantum Newton's cradle has served as a paradigmatic testbed for GHD and QGHD, including early studies of the Lieb-Liniger model~\cite{Caux_Doyon_Dubail_Konik_Yoshimura_19,Bastianello2020therm}, experimental tests of GHD~\cite{Schemmer_Bouchoule_Doyon_Dubail_19}, extensions to multicomponent integrable systems~\cite{Scopa2022}, and analyses with QGHD~\cite{Ruggiero_Calabrese_Doyon_Dubail_20,Ruggiero_Calabrese_Doyon_Dubail_21,Scopa_2023}.\\

We now specialize our result for the density-density correlations to this specific trap quench. The system is initially prepared in a thermal state with double-well potential
\begin{equation}\label{eq:double_well}
    V_0(x) = a_4 x^4 - a_2 x^2, \qquad a_4, a_2 > 0;
\end{equation}
and subsequently quenched to harmonic confinement
\begin{equation}\label{eq:harmonic_post_quench}
    V_1(x) = \frac{\omega^2 x^2}{2}.
\end{equation}
For this protocol, the post-quench dynamics of density-density correlations from the inhomogeneous ground state was derived in Ref.~\cite{Ruggiero_Calabrese_Doyon_Dubail_21}. Here, we extend the analysis to finite temperature. The background dynamics \eqref{eq:hamilton_eom} with harmonic potential can be solved exactly. It corresponds to a rotation of the initial contour $\Gamma_0$ in phase space with period $t_\star\equiv 2\pi/\omega$, namely,
\begin{equation}\label{eq:rotation}
    \mqty(x_t\\ p_t) = \mqty(\cos\omega t & \sin\omega t/\omega \\ -\omega\sin\omega t & \cos\omega t)\mqty(x_0\\ p_0) \, 
\end{equation}
which can be used to evolve the initial phase-space occupation \eqref{eq:FD-distr} also at finite temperatures.\\
Since $s$ is a co-moving coordinate, its value in the final QGHD correlations is determined by the initial phase-space point $(x_0,p_0)$ that evolves to position $x$ at time $t$. For a given $(x,t)$, inverting the forward map~\eqref{eq:rotation} generically yields multiple solutions $x_0^{(a)}(x,t)$, one for each local Fermi point $a$ at that position. 

At $t=0$, there are exactly two Fermi points for every $x\in[-R,R]$, corresponding to the two momentum branches $p_\pm(x)\equiv \pm p_F(x)$ (cf. Eq.~\eqref{eq:fermi_point}). Each branch contributes a single solution $x_0$ for a given $x$. Under the phase-space rotation, however, 
several distinct initial positions \(x_0\) on a given momentum branch can be mapped to the same final position $x$, thereby increasing the number of local Fermi points beyond two and producing split Fermi seas. The total number of solutions $x_0^{(a)}(x,t)$ therefore coincides with the number of local Fermi points at $(x,t)$. For $V_0(x)$ in Eq.~\eqref{eq:double_well}, one finds up to four distinct solutions generated during the dynamics.

To distinguish the two branches of the initial contour, we define 
\begin{equation}\label{eq:s_coordinate_time}
    s_+(x_0)=\frac{\pi}{\T}\int_{-R}^{x_0}\frac{\dd y}{p(y)},
    \qquad
    s_-(x_0)=2\pi-s_+(x_0).
\end{equation}
If the $a$-th Fermi point originates from the branch $p_{\sigma_a}(x_0)=\sigma_a p(x_0)$, with $\sigma_a=\pm1$, its co-moving coordinate is
\begin{equation}
    s_a(x;t)=s_{\sigma_a}\big(x_0^{(a)}(x,t)\big).
\end{equation}
The corresponding initial position is determined by
\begin{equation}
    x=x_0^{(a)}\cos\omega t
    +\sigma_a p\big(x_0^{(a)}\big)\frac{\sin\omega t}{\omega}.
\end{equation}
Consequently, the Jacobian entering the density-density correlation formula~\eqref{eq:dens-dens-chiral-boson} is
\begin{equation}\label{eq:jacobian_double_well}
    \abs{\dv{s_a(x;t)}{x}}
    =
    \frac{\pi}{\T}\frac{1}{p\big(x_0^{(a)}\big)}
    \abs{\dv{x_0^{(a)}}{x}}
    =
    \frac{\pi}{\T}\frac{1}{p\big(x_0^{(a)}\big)}
    \frac{1}{
    \abs{\cos\omega t
    +\sigma_a p'\big(x_0^{(a)}\big)\sin\omega t/\omega}}.
\end{equation}
By inserting the expressions for $s_a(x;t)$, the associated Jacobians $|\dd s_a/\dd x|$, and the finite-temperature CFT correlation \eqref{eq:dphidphi_torus}, in the QGHD prediction \eqref{eq:dens-dens-chiral-boson}, we arrive at a closed analytical expression for $\langle\delta\hat\rho(x)\delta\hat\rho(y)\rangle_{t;\beta}$.\\

Before moving to the comparison with numerics, we briefly discuss the expected regime of validity of Eq.~\eqref{eq:dens-dens-chiral-boson}. The finite-temperature QGHD framework developed in Secs.~\ref{sec:qghd} and~\ref{sec:luttinger} identifies the dimensionless inverse-temperature scale of the problem as $\beta_\text{eff}\propto \beta/{\T}$ (cf. \eqref{eq:torusBC}-\eqref{eq:Omega}). For $\beta/{\T}\gg 1$, thermal effects are therefore negligible, and we expect our results to agree with the zero-temperature prediction of Ref.~\cite{Ruggiero_Calabrese_Doyon_Dubail_21}. Conversely, for $\beta/{\T}\ll 1$, 
thermal broadening around the zero-temperature Fermi contour becomes substantial, so that the local Fermi points underlying the low-energy TLL description are no longer sharply defined. The zero-temperature contour can then no longer be regarded as a weakly fluctuating reference background, and Eq.~\eqref{eq:dens-dens-chiral-boson} is expected to lose accuracy.
In the intermediate regime $\beta/{\T}=O(1)$, on the other hand, Eq.~\eqref{eq:dens-dens-chiral-boson} should display appreciable deviations from the zero-temperature result of Ref.~\cite{Ruggiero_Calabrese_Doyon_Dubail_21}, while remaining predictive when compared with exact numerical data. We mention that the regime $\beta/{\T}=O(1)$ is compatible with temperatures typically achieved in modern cold-atom platforms. In Ref.~\cite{Takacs_Zhang_Calabrese_Dubail_Rigol_Scopa_25}, finite-temperature inhomogeneous TLL predictions for one-body correlations and the corresponding momentum distribution were shown to agree with both quantum Monte Carlo data and experimental measurements in optical-lattice setups.\\

\begin{figure}[t!]
    \centering
    \includegraphics[width=\textwidth]{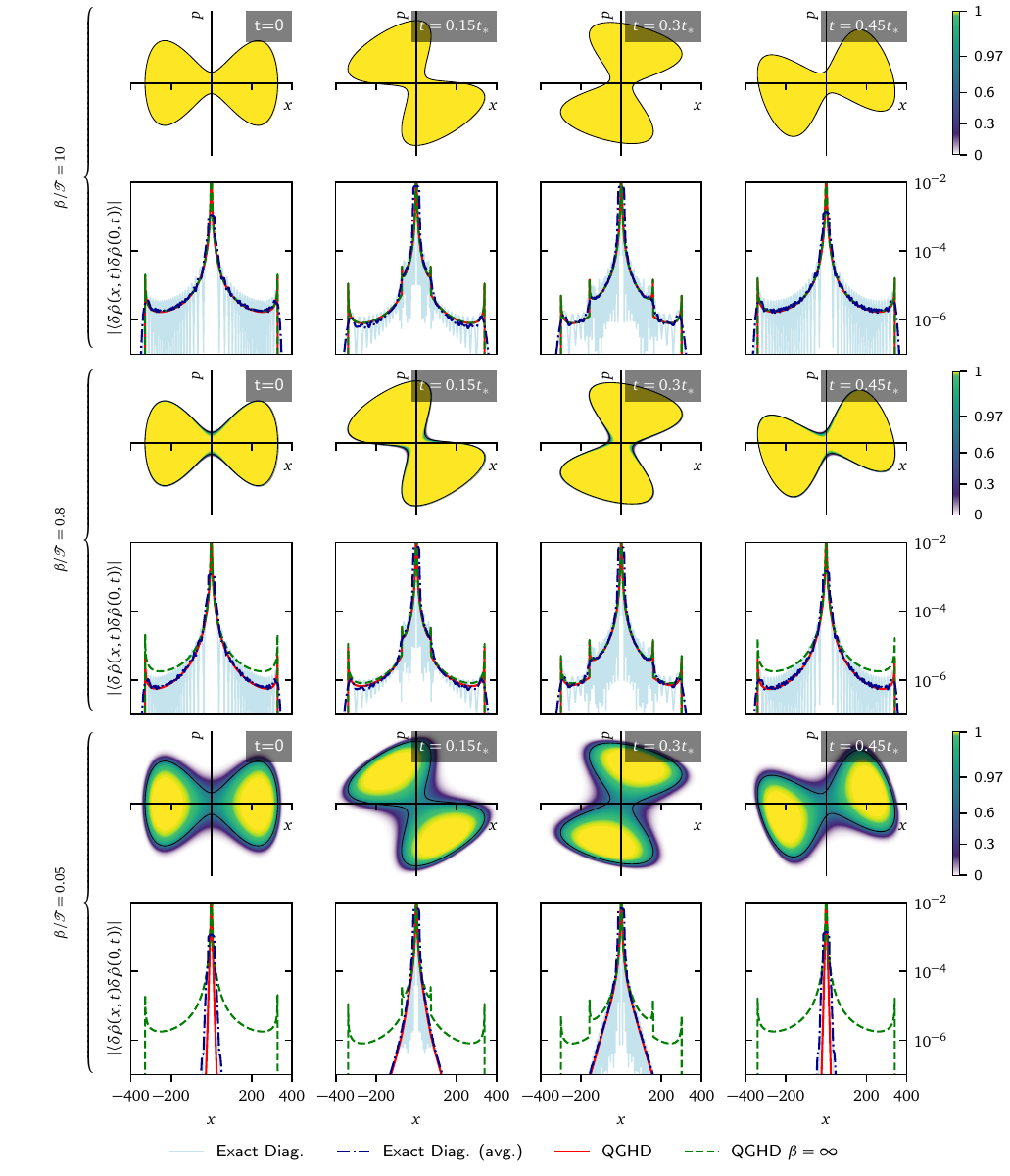}
    \caption{Occupation function and density-density correlations after a double-well to harmonic quench for three temperature regimes. Each two-row block shows the occupation function $n_{t;\beta}(x,p)$ in phase space (upper row) and the corresponding density-density correlation on a logarithmic scale (lower row) at different times. The exact lattice results are shown in light blue and with their short-wavelength oscillations averaged out in dark blue. These are compared with the QGHD predictions at zero temperature (green) and finite temperature (red).
        Low-temperature regime,~$\beta/\T=10$: the zero-temperature and finite-temperature curves are indistinguishable from each other.
        Intermediate regime,~$\beta/\T=0.8$: thermal effects are visible at $t=0$ but are suppressed as the occupation function rotates. 
        High-temperature regime,~$\beta/\T=0.05$: the initial thermal deviation is substantial, yet correlations are dynamically enhanced toward the zero-temperature values during the evolution. The parameters used in the exact numerical data are as follows: system size $L=801$; number of particles $N=44$; pre-quench potential parameters $a_4=6/L^4$, $a_2=1/L^2$; post-quench potential parameter $\omega = 1/L$.
    }
    \label{fig:harmonic_trap_plots}
\end{figure}

In Fig.~\ref{fig:harmonic_trap_plots} we compare the result in Eq.~\eqref{eq:dens-dens-chiral-boson} with both the exact numerical data and the zero-temperature result of Ref.~\cite{Ruggiero_Calabrese_Doyon_Dubail_21} across the three temperature regimes. See Appendix~\ref{app:num} for details on the numerical implementation. In the low-temperature regime ($\beta/\T=10$), the correlation dynamics from the zero-temperature and finite-temperature equilibrium initial state are, as expected, indistinguishable. In the intermediate regime ($\beta/\T=0.8$), thermal effects produce a visible suppression of the correlation in the initial state. However, during the post-quench evolution, the deviation from the ground-state evolution becomes progressively smaller, upon overlapping with the zero-temperature curve up to times $t\simeq 0.5 t_\star$. From there on, the TG dynamics is periodic and reproduces the thermal effects that were present in the initial state. Such correlation enhancement during the dynamics is even more pronounced in the high-temperature regime ($\beta/\T=0.05$), where the initial-state correlation is substantially different from that in the ground state. Moreover, and again as expected, in this regime Eq.~\eqref{eq:dens-dens-chiral-boson} shows visible deviations from the numerical data at $t=0$. Yet, during the dynamics, such correlations are enhanced toward their corresponding low- and zero-temperature predictions. We thus observe that, out of equilibrium, the finite-temperature QGHD result is actually predictive much beyond the expected regime of validity.\\

\begin{figure}
    \centering
\includegraphics[width=0.85\textwidth]{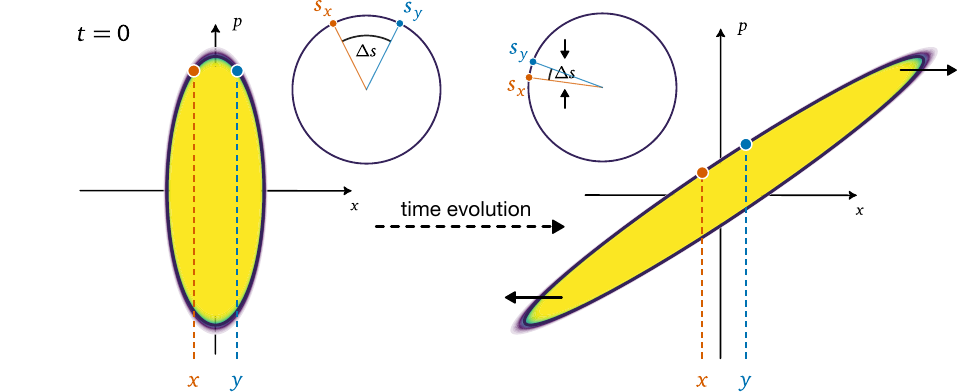}
    \caption{Illustration of the correlation-enhancement mechanism. We consider the expansion of the phase-space occupation $n_t(x,p)$ shown on the left, and focus on two fixed spatial positions $x$ and $y$, at which we wish to characterize the density correlations. Such real-space positions are mapped onto a unit circle in terms of contour coordinates, $s_x$ and $s_y$ respectively for the upper branch. During the evolution, the contour separation $\Delta s$ decreases with time. As a result, the bulk correlations become asymptotically sensitive only to the short-distance sector of the CFT.}
    \label{fig:illustration}
\end{figure}

The reduction of thermal effects visible in Fig.~\ref{fig:harmonic_trap_plots} has a simple interpretation within Eq.~\eqref{eq:dens-dens-chiral-boson}. Indeed, under the post-quench dynamics, the co-moving coordinate $s(x;t)$ assigned to a fixed physical point $x$ changes over time. During the expansion stage of the harmonic evolution, the spatial region occupied by the gas increases and, given two insertion points in real space in the bulk in Fig.~\ref{fig:harmonic_trap_plots}, the corresponding separation $\Delta s_{ab}=s_a(x;t)-s_b(y;t)$ between the relevant insertion points decreases in conformal distance units. As discussed in the following section, the chiral-current correlator entering Eq.~\eqref{eq:dens-dens-chiral-boson} behaves at short contour distance as
\be \label{chiral_corr_short_distance}
\langle\partial\phi(s_a)\partial\phi(s_b)\rangle_{\text{CFT};\beta}\sim \frac{-1}{\Delta s_{ab}^2}+O(1),
\ee
with a leading singularity that is temperature independent, while the regular contribution stays $O(1)$ (as $\Delta s_{ab}\to 0$, the former necessarily dominates over the latter). Converting this into the physical density-density correlation~\eqref{eq:dens-dens-chiral-boson}, however, requires the Jacobian factors $J(s_a)J(s_b)$ that also multiply both terms on the rhs of~\eqref{chiral_corr_short_distance},
\be
J(s_a)J(s_b)\langle\partial\phi(s_a)\partial\phi(s_b)\rangle_{\text{CFT};\beta}\sim J(s_a)J(s_b)\left[\frac{-1}{\Delta s_{ab}^2}+O(1)\right],
\ee
and ultimately, in the regime relevant to Fig.~\ref{fig:harmonic_trap_plots}, this reproduces the universal equilibrium form $\sim 1/(x-y)^2$ characteristic of TLL theory at equilibrium zero-temperature~\cite{Cazalilla_04,Giamarchi_03}. An illustration of this enhancement mechanism is given in Fig.~\ref{fig:illustration}. 

Note, however, that the harmonic evolution is periodic, and so this reduction of thermal effects can only be transient, reversed during the subsequent contraction stage. In the following section, instead, we focus on a pure expansion protocol, for which the role of the Jacobians can be made precise and the enhancement mechanism established quantitatively as a genuine long-time limit. \\
\section{Dynamical enhancement of correlations after trap release}
\label{sec:trap_release}
\begin{figure}[t!]
 \centering
 \includegraphics[width = \linewidth]{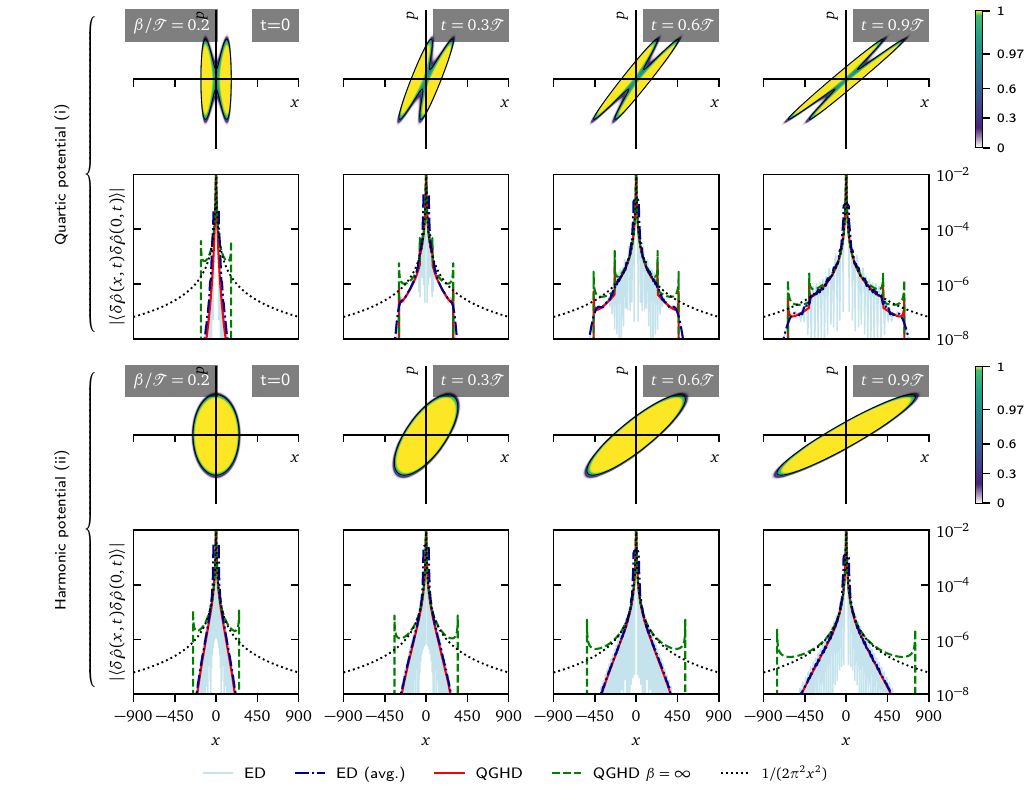}
 \caption{Trap release in the finite-temperature regime ($\beta/\T=0.2$) from two pre-quench potentials: (i)~a double-well quartic potential, $V_0=a_4x^4-a_2x^2$, with system size $L=1801$; number of particles $N=22$; pre-quench parameters $a_2=1/\ell^2$, $a_4=6/\ell^4$, $\ell=401$; and (ii)~a harmonic potential, $V_0=a_2x^2$, with system size $L=1801$; number of particles $N=50$; pre-quench parameter $a_2=1/\ell^2$, $\ell=901$. The parameters are chosen such that the gas remains within the lattice throughout the expansion, ensuring that finite-size effects do not affect the results. Each two-row block shows the occupation function $n_{t;\beta}(x,p)$ in phase space (upper row) and the corresponding density-density correlation on a logarithmic scale (lower row) at times $t=0,\,0.3,\,0.6,\,0.9$ in units of $\T$, see Eq.~\eqref{eq:s_mapping}, with $\T$ the time required for modes to travel from one edge of the cloud to the other. In the phase-space plots, the black solid contour marks the boundary of the zero-temperature state. The exact lattice result is shown in light blue, and with its short-wavelength oscillations averaged out in dark blue. The lattice results are compared with the QGHD predictions at zero temperature (green) and finite temperature (red). The black dotted line is the equilibrium TLL result for the density-density correlation at zero temperature, $1/(2\pi^2 x^2)$, toward which the out-of-equilibrium correlations relax at late times. In the double-well case, thermal effects and split-Fermi-sea structures produce large deviations from this curve at early times, while the harmonic case develops no split Fermi seas; in both cases, as the gas expands, these deviations are progressively suppressed and the correlations are enhanced toward their zero-temperature equilibrium form.} 
 \label{fig:trap_release}
\end{figure}

We thus turn to a trap release protocol, $V_1(x)=0$. In contrast with the harmonic
quench, the expansion is non-periodic and permits an explicit asymptotic
calculation.
For the trap release, the phase-space dynamics \eqref{eq:hamilton_eom} reduces to a symplectic shear,
\begin{equation}\label{eq:symplectic_shear}
    \mqty(x_t\\ p_t) = \mqty(1 & t\\ 0 & 1)\mqty(x_0\\p_0).
\end{equation}
This may be viewed as the limiting case $\omega\to 0$ of the harmonic-trap dynamics in Eq.~\eqref{eq:rotation}.
In Fig.~\ref{fig:trap_release}, we show the results for the trap release in the finite-temperature regime $\beta/{\T}=0.2$, comparing two representative pre-quench potentials: the double-well potential~(i) and a harmonic potential~(ii). 
For the double-well case, the early-stage dynamics is characterized by both visible thermal effects and the emergence of split Fermi-sea structures. Hence, the correlations differ significantly from both the zero-temperature prediction and the equilibrium result. At large times, however, the enhancement mechanism takes place. The resulting phase-space distribution of quasiparticles increasingly resembles that of a connected, boosted Fermi sea. The split-Fermi-sea structure, typical of genuinely out-of-equilibrium configurations, becomes less pronounced in the bulk, 
and the particle occupation progressively approaches, in shape, that of a single connected Fermi sea in the bulk up to a momentum boost that does not affect the density correlations because of Galilean invariance.
For the harmonic pre-quench potential, the initial Fermi contour is an ellipse and no split Fermi-sea structures emerge during the expansion: each spatial point retains exactly two local Fermi points at all times. Nonetheless, the enhancement mechanism operates in the same way. The correlations, initially suppressed by temperature, are progressively enhanced toward their zero-temperature equilibrium form as the gas expands. 
In both protocols, this simplification of the occupation function's shape,
however, does not by itself explain the convergence observed in Fig.~\ref{fig:trap_release}. While it offers a qualitative, semiclassical picture of the approach to homogeneous equilibrium, a quantitative account of the suppression of finite-temperature effects at large times requires combining the exact ballistic evolution of the Fermi contour (encoded in the time-dependent map $x\mapsto s(x;t)$) with the underlying conformal structure of the density-density correlations.
  \\

We therefore consider the large-time asymptotics of the result in Eq.~\eqref{eq:dens-dens-chiral-boson}. We proceed as follows. We take the initial potential $V_0(x)$ to be a generic even function, locally invertible near the edges $x=\pm R$, and satisfying $V_0(\pm R)=\mu$ and $V_0'(\pm R)\neq 0$. For the trap-release dynamics~\eqref{eq:symplectic_shear}, the Jacobians entering Eq.~\eqref{eq:dens-dens-chiral-boson} are then given by
\begin{equation} \label{Jacobians}
   J(s_a)\equiv \dv{s_a(x;t)}{x}
    =
    \frac{\pi}{\T}
    \frac{1}{\pm p(x_0^{(a)})-tV_0'(x_0^{(a)})}
    \overset{t\to\infty}{\sim}\frac{1}{t}.
\end{equation}
The product of the Jacobians contributes to Eq.~\eqref{eq:dens-dens-chiral-boson} with an overall factor $O(t^{-2})$. 

We next consider the large-time behaviour of the correlation $\langle \de\phi(s)\de\phi(s')\rangle_{\text{CFT};\beta}$. To determine its asymptotics, we need to track the scaling of $x_0^{(a)}(x;t)$ and of the associated co-moving coordinate $s_a(x;t)$. From Eq.~\eqref{eq:symplectic_shear}, one simply finds
\begin{equation}
    \pm p(x_0^{(a)})=\frac{x-x_0^{(a)}}{t},
\end{equation}
where the sign distinguishes the two branches of the initial Fermi contour. Since $x_0^{(a)}$ remains bounded within the initial trap, one has $\lim_{t\to\infty}p(x_0^{(a)})=0$. The relevant initial coordinates therefore approach the edges, $x_0^{(a)}\to\pm R$, in the large-time limit.
Without loss of generality, we focus on the branch $x_0\to-R$, and thus suppress the superscript $(a)$ in the expression below. In Appendix~\ref{app:trap_release_asymptotics}, we show that $x_0(x;t)$ admits the expansion
\begin{equation}\label{eq:original_point}
    x_0(x;t)
    =
    -R-\frac{g_1}{2}\rbracket{\frac{x+R}{t}}^2
    +O(t^{-4}),
\end{equation}
where $g_1=1/V_0'(-R)$.
This shows that the dominant contribution at large times originates from $x_0\to -R$, the edge of the gas, corresponding to slow modes with velocity $v\equiv p\to 0$.

We now substitute Eq.~\eqref{eq:original_point} into Eq.~\eqref{eq:s_coordinate_time} and expand around $x_0\approx-R$ to obtain
\begin{equation}
    s(x,t)
    =
    \frac{\pi|g_1|}{\T}\frac{x+R}{t}
    +O(t^{-3}).
\end{equation}
Hence, at late times, $s(x;t)=O(t^{-1})$; see Appendix~\ref{app:trap_release_asymptotics} for details. More precisely, for two points $x$ and $y$ whose Fermi points originate from the same edge,
\begin{equation}\label{eq:s_difference_long_time}
    s(x;t)-s(y;t)
    =
    \frac{\pi|g_1|}{\T}\frac{x-y}{t}
    +O(t^{-3}).
\end{equation}
An analogous relation holds for the contributions originating from the opposite edge. By contrast, Fermi points originating from different edges remain separated by an $O(1)$ distance along the contour. 

Starting from the zero-temperature limit, the short-distance CFT correlator in Eq.~\eqref{eq:dphidphi_torus} reduces to~\cite{DiFrancesco_CFT}
\begin{equation}
    \expval{\partial\phi(z)\partial\phi(0)}_{\text{CFT};\beta\to\infty}
    \overset{z \to 0}{\simeq}
    \frac{-1}{4\sin^2(z/2)}
    \overset{z \to 0}{\simeq}
    -\frac{1}{z^2}.
\end{equation}
Combining this short-distance behaviour with Eq.~\eqref{eq:s_difference_long_time} and with the Jacobians~\eqref{Jacobians}, the contribution associated with each edge becomes 
\begin{equation}
    \frac{1}{4\pi^2}
    \left|J\big(s(x;t)\big)\right|\, \left|J\big(s(y;t)\big)\right|
    \expval{
        \partial\phi\big(s(x;t)\big)
        \partial\phi\big(s(y;t)\big)
    }_{\text{CFT};\beta\to\infty}
    \overset{t\to\infty}{\longrightarrow}
    -\frac{1}{4\pi^2(x-y)^2}.
\end{equation}
Since the two edges give equal contributions, while the terms connecting opposite edges vanish as $O(t^{-2})$, one obtains, for $x\neq y$,
\begin{equation}\label{eq:long-time-equilibrium-correlation}
    \lim_{t\to\infty}
    \expval{\delta\hat\rho(x)\delta\hat\rho(y)}_{t;\beta=\infty}
    =
    \frac{-1}{2\pi^2(x-y)^2}.
\end{equation}
This is precisely the long-wavelength homogeneous density-density correlation for a connected Fermi sea~\cite{Giamarchi_03,Cazalilla_04}. 

The same conclusion holds at finite temperature. Indeed, 
chiral-current correlators are now given by \eqref{eq:dphidphi_torus}, and, 
from the expansion of the Jacobi theta function around the origin,
\begin{equation}
    \theta_1(z/2|\Omega)
    =
    \frac{z}{2}\theta_1'(0|\Omega)+O(z^3),
\end{equation}
one finds
\begin{equation}\label{eq:finite_temperature_small_argument}
    \partial_z^2\log\theta_1(z/2|\Omega)
    \overset{z \to 0}{\simeq}
    -\frac{1}{z^2}.
\end{equation}
The leading singularity is therefore independent of temperature and coincides with the zero temperature result. All temperature-dependent regular terms, including the contribution of the charge zero mode, are multiplied by the $O(t^{-2})$ Jacobian factor and vanish in the large-time limit. Consequently,
\begin{equation} \label{rhorho_tlimit}
    \lim_{t\to\infty}
    \expval{\delta\hat\rho(x)\delta\hat\rho(y)}_{t;\beta}
    =
    -\frac{1}{2\pi^2(x-y)^2},
\end{equation}
independently of the initial temperature $\beta$. 
Eq.~\eqref{rhorho_tlimit} should be understood within the hydrodynamic scaling regime. Although $\Delta s\sim(x-y)/t$ tends to zero, the physical distance $|x-y|$ is not taken to zero. It remains large compared with the local microscopic scale, while the expansion makes the corresponding contour separation small compared with the thermal and inhomogeneity scales. Thus, the short-distance limit concerns the contour coordinate $s$ and does not conflict with the long-wavelength nature of the standard TLL description.

The above argument shows that, under mild assumptions on the pre-quench confinement $V_0$, the finite-temperature density-density correlations asymptotically approach the corresponding zero-temperature equilibrium result for any bulk positions $x$ and $y$ fixed.
Note that the same vanishing Jacobian also controls the thermal broadening of the occupation around each local Fermi edge; see Appendix~\ref{app:enhancement-mechanism} for details. In particular, during the expansion the associated thermal width does not get dynamically amplified. This is consistent with the effective description remaining accurate at late times. 
\\

We also recall, as visible from Fig.~\ref{fig:trap_release}, that finite-temperature QGHD captures the transient dynamics toward this large-time asymptotic result. Within the QGHD framework developed here, this crossover is controlled by the ratio between the contour separation $|\Delta s|$ and the thermal scale $\beta_\text{eff}$. As the temperature increases, shorter contour separations are required for the temperature-independent singularity to dominate. Since $\Delta s\sim t^{-1}$ after trap release, the onset of the asymptotic regime is therefore shifted to later times at higher temperatures. If the initial temperature lies within the TLL regime, namely if $\beta/{\T}=O(1)$ or larger, our finite-temperature QGHD prediction remains accurate throughout the dynamics since the chiral-current correlator in the co-moving coordinate does not evolve. At higher initial temperatures, namely when $\beta/{\T}\ll 1$, it becomes accurate only at large times, following the onset of the aforementioned enhancement mechanism. Notice, however, that this approach to the short-distance sector of the CFT occurs while the gas becomes increasingly dilute in real space.

{We conclude by noting that the correlation enhancement discussed here is closely related to the mechanism exploited in delta-kick cooling of expanding quantum gases~\cite{Ammann_Christensen_97,Myrskog_Fox_Moon_Kim_Steinberg_00,dupays2021delta,Dupays_Yang_delCampo_23}. In such protocols, the gas is first allowed to expand freely, so that particles with different momenta progressively separate in space. A subsequent short trapping pulse then provides a position-dependent momentum kick that compensates the expansion and strongly reduces the momentum spread of the cloud~\cite{Ammann_Christensen_97,Myrskog_Fox_Moon_Kim_Steinberg_00,dupays2021delta,Dupays_Yang_delCampo_23}. In our setting, although one could in principle perform the same additional kick, we consider only the expansion stage, since we aim to characterize the quench dynamics following the trap release itself. The resulting state should therefore not be interpreted as a cooled thermal state, but rather as a boosted Fermi sea. Nevertheless, at the level of density correlations, the expansion alone is sufficient to suppress the thermal contributions in the bulk and drive the correlations toward their zero-temperature equilibrium form.
}

\section{Conclusion}\label{sec:conclusion}
In this work, we studied the nonequilibrium dynamics of connected density-density correlations at finite temperature following a trap quench. Our analysis relies on the QGHD framework, which we extend here to thermal initial states. Focusing on the TG limit of the Lieb-Liniger gas allows us to obtain closed analytical results by exploiting the conformal structure of the corresponding low-energy theory. We benchmark these predictions against exact free-fermion numerics over a range of temperatures. While the finite-temperature QGHD framework applies to generic trap quenches, we illustrate its predictions using two representative protocols: a double-well-to-harmonic quench and a trap release. In both cases, we find that finite-temperature QGHD remains quantitatively accurate throughout the expected low-temperature regime and, remarkably, can remain predictive even beyond it during the nonequilibrium evolution.
In particular, we observe that the post-quench dynamics can suppress thermal effects in the correlations, bringing them progressively closer to their zero-temperature values. Such a dynamical enhancement was first identified in Ref.~\cite{DeAngelis_DeNardis_Scopa_24} for a domain-wall melting protocol. Here, we show that a similar phenomenon occurs for inhomogeneous trap quenches and provide an analytical characterization of its long-time behaviour. 
For a trap-release protocol, we show that, at long times, the relevant insertion points approach the edges of the initial Fermi contour and their separations in conformal coordinates shrink as $t^{-1}$, compensated by an equally vanishing Jacobian. This feature is a direct consequence of the expansion and of the underlying conformal invariance of the TG gas. 
As a result, the density-density correlations approach the zero-temperature equilibrium form $\sim1/x^2$, independently of the initial temperature.
\\ 

A natural extension of this work is to move beyond the TG limit and investigate the effect of finite interactions. Due to the universal character of TLL theory, we expect the dynamical suppression of thermal effects to persist also away from the TG limit. It remains an open question, however, to understand quantitatively how this correlation enhancement is modified by the quasiparticle dressing, which affects both the hydrodynamic background and the effective theory of fluctuations.

Finally, we note that the evolution equation for the Wigner function~\eqref{eq:Wigner-evo} corresponds to the leading, first-order term in the Moyal expansion~\cite{Moyal_49}. For the particular quenches in Sec.~\ref{sec:double_well_results} and Sec.~\ref{sec:trap_release}, this equation is exact due to the specific form of the potential function $V_1$. For generic potentials, however, higher-order terms in the Moyal expansion provide additional quantum corrections~\cite{Fagotti_17,Fagotti_20,urilyon2024quantum}. It would therefore be interesting to investigate how these corrections modify the QGHD description and how they interplay with both thermal and quantum fluctuations.

\ack{GL would like to thank Léonce Dupays for insightful discussions. SS and PR acknowledge Attila Takács, 
Pasquale Calabrese, Jérôme Dubail and  Benjamin Doyon for discussions and previous collaboration on related topics. SS is supported by the MSCA Grant No.~101103348 (GENESYS). 
PR acknowledges support from the UK Engineering
and Physical Sciences Research Council (EPSRC) through a
New Investigator Award, grant number EP/Y015363/1.
Views and opinions expressed are those of the authors only and do not necessarily reflect those of the European Union or the European Research Council Executive Agency. Neither the European Union nor the granting authority can be held responsible for them.
}

\appendix
\section{Numerical implementation}\label{app:num}
In this appendix we describe the exact numerical scheme used to benchmark our analytical predictions. The computation is standard and relies on the fact that, for density correlations, the TG limit is equivalent to non-interacting spinless fermions. As such, the entire many-body problem can be reduced to a single-particle one and can be solved by diagonalizing an $L\times L$ matrix, where $L$ denotes the system size in units of lattice spacing.
\subsection{Lattice discretization}
We discretize the continuum on a uniform lattice of $L$ sites, with site $i$ located at $x_i=i-L/2$, so that the trap is centered on the lattice. The kinetic term is obtained by replacing the second derivative with a finite-difference discretization scheme. The resulting Hamiltonian at time $t$ reads
\begin{equation}\label{eq:app-lattice-H}
    \hat H(t) = \sum_{i=1}^{L}\Big[
        \rbracket{J_0+V_i(t)}\hat c_i^\dagger \hat c_i
        +J_1\,\rbracket{\hat c_i^\dagger \hat c_{i+1}+\hat c_{i+1}^\dagger \hat c_i}+
        J_2\,\rbracket{\hat c_i^\dagger \hat c_{i+2}+\hat c_{i+2}^\dagger \hat c_i}
    \Big],
\end{equation}
where $\hat c_i,\hat c_i^\dagger$ are spinless fermionic operators obeying $\{\hat c_i,\hat c^\dagger_j\}=\delta_{ij}$, and the hopping amplitudes
\begin{equation}
    J_0=\frac{5}{4},\qquad J_1=-\frac{2}{3},\qquad J_2=\frac{1}{24}
\end{equation}
are (half) the fourth-order finite-difference coefficients of the second derivative~\cite{FD_coefficients}. We impose {open} boundary conditions and take the box large enough such that finite size effects are negligible. $V_i(t)=V(x_i;t)$ denotes the onsite potential and realizes the quench protocol of the main text,
\begin{equation}
    V_i(t) =\begin{cases}
        a_4\,x_i^{4}-a_2\,x_i^{2}, & t=0 \quad\text{(double well)},\\[4pt]
        \tfrac12\,\omega^2\,x_i^{2}, & t>0 \quad\text{(quench to harmonic trap)},\\[4pt]
        0, & t>0 \quad\text{(trap release)}.
    \end{cases}
\end{equation}

\subsection{Initial thermal two-point correlation function}
In this section we explain how the two-point function is computed for thermal states. Since the initial state is thermal, it obeys Wick's theorem~\cite{Gaudin_59}, and each observable can be decomposed in terms of two-point correlators. We first diagonalize the pre-quench single-particle Hamiltonian $\hat H(0)=\sum_{i,j} \hat c^\dagger_i\, h_0 \, \hat c_j$, namely $h_0\ket{v_k}=E_k\ket{v_k}$. The two-point correlation function then reads
\begin{equation}\label{eq:app-C0}
    C_{ij}(0)=\expval{\hat c_i^\dagger \hat c_j}_{0;\beta}
    =\sum_{k=1}^{L}\braket{v_k}{i}\, n_F(E_k)\,\braket{j}{v_k},
    \qquad
    n_F(E)=\frac{1}{1+e^{\beta(E-\mu)}},
\end{equation}
where $\ket{j}$ denotes the real-space (site) basis, and $n_F$ is the Fermi-Dirac occupation. At each inverse temperature $\beta$, the chemical potential $\mu=\mu(\beta)$ is fixed by requiring the mean particle number $N=\sum_k n_F(E_k)$ to be the same at all temperatures considered, so that comparisons are made at fixed density.
\subsection{Time evolution and correlations}
For $t>0$ the state evolves unitarily with the post-quench Hamiltonian $\hat H_1\equiv \hat H(t>0)$. Since $\hat H_1$ is quadratic, the state remains Gaussian and the correlation matrix simply evolves as
\begin{equation}\label{eq:app-Ct}
    C(t)=e^{i h_1 t}\,C(0)\,e^{-i h_1 t},
\end{equation}
with $\hat H_1=\sum_{i,j} \hat c^\dagger_i \, h_1\, \hat c_j$, and can be evaluated by diagonalizing $h_1$. The connected density-density correlation is then evaluated in terms of $C(t)$ using Wick's theorem. For $i\neq j$, one finds
\begin{equation}\label{eq:app-dd}
    \expval{\delta\hat\rho_i(t)\,\delta\hat\rho_j(t)}=-|C_{ij}(t)|^2.
\end{equation}
Eq.~\eqref{eq:app-dd} is shown in Figures~\ref{fig:harmonic_trap_plots} and \ref{fig:trap_release} as a function of the site positions $x_i,x_j$, and compared to the QGHD prediction of Eq.~\eqref{eq:dens-dens-chiral-boson}.
Convergence to the continuum is controlled by increasing the number of lattice
sites at fixed particle number. In the figures, the short-wavelength
oscillations are smoothed by a local average over a few sites,
\begin{equation}
    \overline{\expval{\delta\hat{\rho}_i(t)\,\delta\hat{\rho}_j(t)}}
    =\frac{1}{\Delta x}\sum_{k=-\Delta x/2}^{\Delta x/2}
      \expval{\delta\hat{\rho}_i(t)\,\delta\hat{\rho}_{j+k}(t)},
\end{equation}
where $\Delta x$ is chosen such that it is greater than the oscillation wavelength
but smaller than the hydrodynamic length scale, so that the oscillations are
averaged out while the coarse-grained profile remains unchanged.
\section{Trap release asymptotic expansion for generic potentials}
\label{app:trap_release_asymptotics}
This appendix details the asymptotic expansion of the coordinate $s(x,t)$ in the long-time limit for a trap release protocol from an arbitrary confining potential $V_0(x)$. 

\subsection{Expansion of the initial coordinate \texorpdfstring{$x_0$}{x\_0}}
For a trap release, the post-quench potential is zero, and the phase-space trajectories are given by $x = x_0+p(x_0)t$, where $p(x_0)=\sqrt{2(\mu-V_0(x_0))}$. Squaring this relation yields
\begin{equation}
    V_0(x_0) = \mu - \frac{(x-x_0)^2}{2t^2}.
\end{equation}
We assume the initial potential satisfies $V_0(\pm R)=\mu$ and $V'_0(\pm R)\neq 0$. These are the only conditions needed for the late-time scaling used in the main text. Without loss of generality, we focus on the limit $t\to \infty$ for the left boundary, $x_0 \to -R$. Using the Lagrange inversion theorem, the inverse function $V^{-1}_0$ around $\mu$ can be expressed as a series:
\begin{equation}
    x_0 = V^{-1}_0\rbracket{\mu-\frac{(x-x_0)^2}{2t^2}} = -R+\sum_{n=1}^{\infty}\frac{g_n(-1)^n}{2^nn!}\rbracket{\frac{x-x_0}{t}}^{2n},
\end{equation}
with the leading coefficient
\begin{equation}
    g_1 = \frac{1}{V'_0(-R)}.
\end{equation}
For a confining potential, $V'_0(-R)<0$, making $g_1<0$. At leading order, $(x-x_0)\approx (x+R)$, so the $n=1$ term gives
\begin{equation}
    x_0 = -R-\frac{g_1}{2}\frac{(x+R)^2}{t^2}+O(t^{-4}).
\end{equation}
\subsection{Evaluation of the co-moving coordinate \texorpdfstring{$s$}{s}}
The co-moving coordinate $s(x,t)$ for the corresponding Fermi point is defined as
\begin{equation}
    s(x,t) = \frac{\pi}{\T}\int_{-R}^{x_0}\frac{\dd y}{p(y)} = \frac{\pi}{\T}\int_{-R}^{-R+\delta}\frac{\dd y}{\sqrt{2(\mu-V_0(y))}},
\end{equation}
where we introduced the displacement $\delta \equiv x_0+R = -\frac{g_1}{2}\frac{(x+R)^2}{t^2}$. By Taylor-expanding the potential near the boundary $y=-R$,
\begin{equation}
    V_0(y) = \mu+V'_0(-R)(y+R)+O((y+R)^2),
\end{equation}
which implies $\mu-V_0(y) = \abs{V'_0(-R)}(y+R)+O((y+R)^2)$. Substituting this into the momentum yields
\begin{equation}
    p(y) = \sqrt{2\abs{V'_0(-R)}(y+R)}\sbracket{1+O(y+R)}.
\end{equation}
We evaluate the integral by substituting $u=y+R$:
\begin{align}
    \int_{-R}^{-R+\delta}\frac{\dd y}{p(y)} &= \int_0^{\delta}\frac{\dd u}{\sqrt{2|V'_0(-R)|u}\sbracket{1+O(u)}}\nonumber\\
                                            &= \frac{2\sqrt{\delta}}{\sqrt{2\abs{V'_0(-R)}}} +O(\delta^{3/2}).
\end{align}
Recalling the definition of $\delta$, we find $\sqrt{\delta} = \frac{x+R}{\sqrt{2\abs{V'_0(-R)}}\,t}$. Therefore, to leading order in $\delta$,
\begin{equation}
    \int_{-R}^{-R+\delta}\frac{\dd y}{p(y)} \approx \frac{2}{\sqrt{2\abs{V'_0(-R)}}}\cdot\frac{x+R}{\sqrt{2\abs{V'_0(-R)}}\,t} = \frac{x+R}{\abs{V'_0(-R)}\,t}.
\end{equation}
Including the prefactor, the coordinate $s$ simplifies to
\begin{equation}\label{eq:s_general_corrected_final}
    s(x,t) = \frac{\pi}{\T}\frac{x+R}{\abs{V'_0(-R)}\,t} + O(t^{-3}) = \frac{\pi\abs{g_1}}{\T}\frac{x+R}{t} + O(t^{-3}).
\end{equation}
This confirms that $s(x,t)=O(1/t)$ at late times for any generic confining potential.

\section{Thermal broadening around the local Fermi edges}
\label{app:enhancement-mechanism}
In this appendix, we formalize the notion of thermal broadening around the local Fermi edges and examine its post-quench dynamics. We show that for the two protocols considered in the main text, the thermal broadening is not dynamically amplified: it decays as $t^{-1}$ during the trap release and remains bounded for the harmonic quench.
Finally, in Subsec.~\ref{app:jacobian_broadening}, we demonstrate that this behaviour has a geometric origin, establishing that the thermal broadening is directly governed by the Jacobian entering Eq.~\eqref{eq:dens-dens-chiral-boson}.
Let the zero-temperature Fermi contour be defined implicitly by
\begin{equation}
    F_0(x_0,p_0) \equiv \frac{p_0^2}{2}+V_0(x_0)-\mu=0.
\end{equation}
The finite-temperature occupation function is locally described by a Fermi-Dirac distribution,
\begin{equation}
    n_{0;\beta}(x_0,p_0) = \frac{1}{1+e^{\beta F_0}}.
\end{equation}
At a fixed position $x_0$ and inverse temperature $\beta$, we define the thermal broadening $\delta p_{0;\beta}$ as the characteristic momentum scale over which the occupation changes appreciably, i.e.,
\begin{equation}
    \beta F_0(x_0,\,p_0+\delta p_{0;\beta})=O(1).
\end{equation}
Linearizing $F_0$ around the zero-temperature contour, where $F_0(x_0,p_0)=0$, 
\begin{equation}\label{eq:thermal_broad_def}
    \delta p_{0;\beta} \sim \frac{1}{\beta\,\partial_{p_0}F_0}.
\end{equation}
\subsection{Thermal broadening in trap release}
We now compute the thermal broadening $\delta p_{t;\beta}$ for the trap release. The phase-space trajectories are $x_t=x_0+p_0t$, $p_t=p_0$ such that 
\begin{equation}
n_t(x,p) = n_0(x_t-p_t t,\,p_t), \qquad F_t(x_t,p_t) = F_0(x_t-p_0t,\,p_0).
\label{eq:D-release-Ft}
\end{equation}
Differentiating with respect to $p_t$ at fixed $x_t$,
\begin{align}
    \partial_{p_t}F_0(x_t-p_0 t;p_0) &= \partial_{p_0}F_0-t\partial_{x_0}F_0\\
                                     &= p_0-tV_0'(x_t-p_0 t)\label{eq:D-release-dpFt}
\end{align}
At a regular point of the initial contour, $V'_0\neq 0$ and Eq.~\eqref{eq:D-release-dpFt} is dominated by the second term at late times, so that
\begin{equation}
    \abs{\delta p_{t;\beta}}\sim \frac{1}{\beta\abs{p_t-tV_0'(x_t-p_0 t)}}\overset{t\to\infty}{\sim} \frac{1}{t}.
\end{equation}
Thus, thermal broadening around each local Fermi edge decreases under free expansion. The local Fermi sea width $p_+-p_-$ decreases with the same scaling, so their ratio remains of the same order as in the initial state. Therefore, if the initial state lies within the TLL regime, the expansion preserves the required separation of scales and the effective description remains controlled at late times.
\subsection{Thermal broadening in the harmonic quench}
We now compute the thermal broadening $\delta p_{t;\beta}$ for a quench to a harmonic potential. The phase-space dynamics is governed by a rotation
\begin{equation}
x_0 = x_t\cos\omega t - \frac{p_t}{\omega}\sin\omega t, \qquad
p_0 = p_t\cos\omega t + \omega x_t \sin\omega t,
\label{eq:D-harmonic-x0p0}
\end{equation}
such that the evolved contour function is $F_t(x_t,p_t)=F_0(x_0,p_0)$ and 
\begin{align}
    \partial_{p_t}F_t(x_t,p_t)&=\partial_{p_0}F_0\times\pdv{p_0}{p_t} +\partial_{x_0}F_0\times\pdv{x_0}{p_t} \\
                              &=p_0\cos\omega t-\frac{V_0'}{\omega}\sin\omega t
\end{align}
Unlike Eq.~\eqref{eq:D-release-dpFt}, this is a periodic function of $t$ and remains generically $O(1)$ bounded away from both zero and infinity except possibly at isolated times where the combination vanishes. Consequently,
\begin{equation}
    \abs{\delta p_{t;\beta}} \sim \frac{1}{\beta\big|\partial_{p_t} F_t(x_t,p_t)\big|} = O(1).
\label{eq:D-harmonic-deltapT}
\end{equation}
The thermal broadening oscillates periodically without any systematic growth.
\subsection{The Jacobian as a measure of thermal broadening}\label{app:jacobian_broadening}
While evaluating $\partial_{p_t} F_t$ directly from the phase-space trajectories provides an intuitive and straightforward approach for the simple post-quench potentials discussed above, we now show that the thermal broadening is proportional to the Jacobian $J = |\dv{s}{x_t}|$. This formulation holds for generic quenches and confining potentials. Furthermore, it establishes a direct connection to the validity of our effective theory, as $J$ is the very same Jacobian entering in Eq.~\eqref{eq:dens-dens-chiral-boson}.

To this end, we introduce the forward map $M$ of the phase-space $(x_0,p_0)\mapsto(x_t,p_t)$ defined by the Hamiltonian flow~\eqref{eq:hamilton_eom},
\begin{equation}
    M \equiv \mqty(\pdv{x_t}{x_0} & \pdv{x_t}{p_0}\\ \pdv{p_t}{x_0} & \pdv{p_t}{p_0}) \eqqcolon \mqty(A & B \\ C & D),
\end{equation}
so that each column of $M$ describes how an infinitesimal displacement in the initial conditions $(x_0,p_0)$ propagates into a displacement in $(x_t, p_t)$. From Liouville's theorem we have that phase-space volume is preserved and as such $\det M=1$~\cite{Arnold_mechanics}. The backward map $M^{-1}$, which maps $(x_t, p_t) \mapsto (x_0, p_0)$, admits a closed-form expression
\begin{equation}
    M^{-1} = \mqty(D & -B \\ -C & A)\equiv \mqty(\pdv{x_0}{x_t} & \pdv{x_0}{p_t}\\ \pdv{p_0}{x_t} & \pdv{p_0}{p_t}).
\end{equation}
Reading off the matrix entries yields the symplectic identities
\begin{equation}\label{eq:identities}
    \pdv{p_0}{p_t}=\pdv{x_t}{x_0},\qquad \pdv{x_0}{p_t}=-\pdv{x_t}{p_0},
\end{equation}
which we exploit below.
We now evaluate $\partial_{p_t}F_t(x_t,p_t)$. Since the Hamiltonian flow transports $\Gamma_0$ to $\Gamma_t$ via $M$, the evolved contour function is simply $F_t = F_0 \circ M^{-1}$, i.e., $F_t(x_t,p_t)=F_0(x_0,p_0)$. Applying the chain rule and using $\partial_{p_0}F_0 = p_0$ and $\partial_{x_0}F_0 = V_0'(x_0)$,
\begin{align}
    \partial_{p_t}F_t(x_t,p_t)&=\partial_{p_t}F_0(x_0,p_0)\nonumber\\
                              &=p_0\pdv{p_0}{p_t}+V_0'(x_0)\pdv{x_0}{p_t}\nonumber\\
                              &=p_0\sbracket{\pdv{p_0}{p_t}-p_0'\pdv{x_0}{p_t}}\nonumber\\
                              &=p_0\sbracket{\pdv{x_t}{x_0}+p'_0\pdv{x_t}{p_0}}\label{eq:hamilton_identity},
\end{align}
where we used $V_0'(x_0)=-p_0\,p_0'$, and applied the symplectic identities~\eqref{eq:identities}. Recognizing that
\begin{equation}
    \dv{x_t}{x_0}=\pdv{x_t}{x_0}+p_0'\pdv{x_t}{p_0},
\end{equation}
we obtain
\begin{equation}
    \partial_{p_t}F_t = p_0\, \dv{x_t}{x_0}.
\end{equation}
The thermal broadening at time $t$ is therefore
\begin{equation}
    \delta p_{t;\beta} \sim \frac{1}{\beta\,\partial_{p_t}F_t}=\frac{1}{\beta}\cdot \frac{1}{p_0}\cdot\dv{x_0}{x_t} = \frac{\T}{\pi\beta}\cdot\underbrace{\frac{\pi}{\T\,p(x_0)}\cdot\dv{x_0}{x_t}}_{=\,J},
\end{equation}
which yields the general relation
\begin{equation}
    \delta p_{t;\beta}\sim \frac{J}{\beta_{\text{eff}}}.
\end{equation}
This shows that the Jacobian $J$ directly controls the thermal width of the occupation around each local Fermi edge in units of the effective inverse temperature $\beta_\text{eff}$.

\bibliographystyle{vancouver-linked}
\bibliography{references_qghd}

\end{document}